\documentclass[12pt]{iopart}
\usepackage[dvipdfmx]{color}
\usepackage[dvipdfmx]{graphicx}
\usepackage{bm}
\usepackage{times} 
\usepackage{cite}

\begin{document}

\title[Dynamical coupling of dilute magnetic impurities with quantum spin liquid state]
{Dynamical coupling of dilute magnetic impurities with quantum spin liquid state
in the \textit{S} $=$ 3/2 dimer compound Ba$_{3}$ZnRu$_{2}$O$_{9}$}

\author{Takafumi D. Yamamoto, Hiroki Taniguchi, and Ichiro Terasaki}

\address{Department of Physics, Nagoya University, Nagoya 464-8602, Japan}

\ead{tdyamamoto@nagoya-u.jp}

\begin{abstract}
We have investigated the dilute magnetic impurity effect on
the magnetic properties of a quantum spin liquid candidate
Ba$_{3}$ZnRu$_{2}$O$_{9}$ and a spin gapped compound Ba$_{3}$CaRu$_{2}$O$_{9}$.
The magnetic ground state of each compound stands
against 2\% substitution of magnetic impurities for Zn or Ca.
We have found that the magnetic response of these impurities,
which behave as paramagnetic spins,
depends on the host materials
and the difference of the two manifests itself in the Weiss temperature,
which can hardly be explained by the dilute magnetic impurities alone
in the case of Ba$_{3}$ZnRu$_{2}$O$_{9}$.
We consider a contribution from the Ru$^{5+}$ ions
which would appear only in the substituted Ba$_{3}$ZnRu$_{2}$O$_{9}$
and discuss a possible physical meaning of the observed Weiss temperature.
\end{abstract}
\noindent{Keywords:}
{quantum spin liquid, magnetic oxide, impurity effects, magnetic susceptibility}

\submitto{\JPCM}

\maketitle

\section{Introduction}
The impurity effect on quantum spin systems have attracted much interest
because it has often provided us the opportunity
to discover novel physical phenomena.
For example, in various spin gapped (SG) systems
such as the Haldane compound
\cite{Azuma-PRB55-1997,Uchiyama-PRL83-1999,Oosawa-PRB66-2002,Waki-PRB73-2006},
the dilute non-magnetic impurities induce
an exotic long-range antiferromagnetic order
which coexists with the non-magnetic gapped state.
In this context, one of intriguing targets is
a quantum spin liquid (QSL) material,
in which the interacting spins fluctuate
down to the absolute zero temperature
due to competing magnetic interactions
and quantum fluctuations\cite{Balents-Nature-2010}.

Recent intensive studies have found out
many candidates for a QSL state
in real organic/inorganic materials
with geometrical frustration of antiferromagnetic spins
\cite{Hiroi-JPSJ70-2001,Shimizu-PRL91-2003,Okamoto-PRL99-2007,Itou-PRB77-2008,Okamoto-JPSJ78-2009}.
On the other hand, a new route to find QSL materials has been opened
since Kitaev proposed a quantum spin model on a honeycomb lattice,
which hosts a novel QSL state\cite{Kitaev-AnnPhys321-2006}.
Such so-called Kitaev spin liquid has been suggested to realize
in strongly spin-orbit coupled Mott insulators\cite{Jackeli-PRL102-2009},
and the experimental explorations on 4d and 5d transition-metal-based compounds actually
have revealed several spin-orbit-driven spin liquid candidates
\cite{Singh-PRB-82-2010,Singh-PRL-2012,Gretarsson-PRL-110-2013,Plumb-PRB-90-2014,
Sears-PRB-91-2015,Johnson-PRB-92-2015,Winter-JPCM29-2017}.
Nowadays the understanding of the physical properties of the QSL materials is
one of the central issues in condensed matter physics.
However, despite large amounts of theoretical and experimental works,
the effect of intentional impurity doping on the systems
has not been much explored except for several cases
\cite{Okamoto-PRL99-2007,Manni-PRB89-2014-1,Manni-PRB89-2014-2,
Sasaki-JPSJ84-2015,Saito-arxiv-2016,Kelley-PRL119-2017}.

\begin{figure}[!t]
\centering
\includegraphics[width=120.0mm,clip]{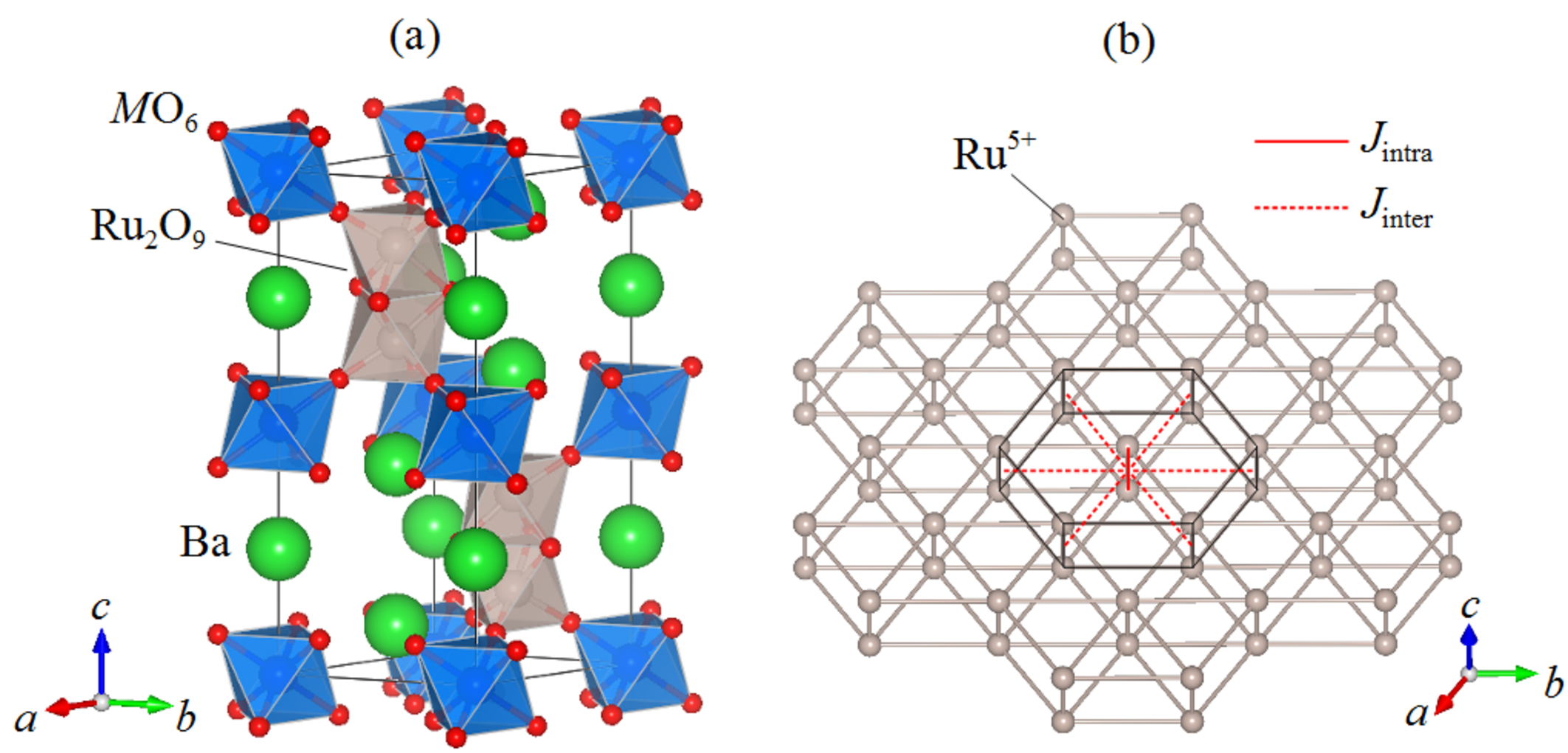}
\caption{(Color Online) (a)Crystal structure of
Ba$_{3}$\textit{M}Ru$_{2}$O$_{9}$
with the space group of \textit{P}6$_{3}$/\textit{mmc},
drawn using VESTA\cite{Momma-VESTA}.
(b) a honeycomb dimer lattice of Ru$^{5+}$ ions
in the \textit{ab} plane layer
with the intra- ($J_{\rm intra}$) and
inter-dimer ($J_{\rm inter}$) exchange coupling.}
\label{fig:structure}
\end{figure}

In this study, we will focus on
a 6H hexagonal perovskite ruthenate Ba$_{3}$ZnRu$_{2}$O$_{9}$,
which is a QSL candidate that we have recently discovered
\cite{Terasaki-JPSJ86-2017}.
This material is a member of the compounds with
the general formula Ba$_{3}$\textit{M}Ru$_{2}$O$_{9}$,
which is composed of the three-dimensional network of
two face-shared RuO$_{6}$ octahedra (Ru$_{2}$O$_{9}$ dimers)
interconnected by the \textit{M}O$_{6}$ octahedra with corner sharing,
as shown in figure \ref{fig:structure}(a).
The \textit{M} site can accommodate various divalent ions,
and the formal valence of Ru is pentavalent in this case.
Then the Ru$^{5+}$ ion (4\textit{d}$^{3}$) is responsible for
the magnetism of the system as a localized moment of $S =$ 3/2.

The magnetic ground state of Ba$_{3}$\textit{M}Ru$_{2}$O$_{9}$
depends on the species of \textit{M}.
For \textit{M} $=$ Co, Ni, and Cu
\cite{Fernandez-JSSC34-1980,Lightfoot-JSSC89-1990,Rijssenbeek-PRB58-1998},
an antiferromagnetic ordering is observed below $T_{\rm N} \sim$ 100 K.
When the \textit{M} sites are occupied by Ca, Sr, and Mg
\cite{Fernandez-JSSC34-1980,Darriet-JSSC19-1976},
paired spins in the Ru$_{2}$O$_{9}$ dimer form a spin singlet,
and then the system is in a non-magnetic SG ground state.
In the case of the \textit{M} $=$ Zn sample,
we have found neither a long-range magnetic order
nor a spin glass transition down to 37 mK
\cite{Terasaki-JPSJ86-2017},
despite the magnetic interaction has been evaluated
to be an energy scale of around 200 K in this family
\cite{Darriet-JSSC19-1976,Senn-JPhys25-2013,Streltsov-PRB88-2013}.
The magnetic susceptibility at low temperatures exhibits
a nearly temperature-independent value of about 10$^{-3}$ emu/mol,
and the specific heat shows a temperature-linear contribution.
These features suggest the presence of a QSL state.
We have proposed that the competition between intra- and inter-dimer interactions
play important role in stabilizing this QSL state.
The exchange coupling pathways are shown in figure \ref{fig:structure}(b). 

We underline several advantages of Ba$_{3}$ZnRu$_{2}$O$_{9}$
in investigating the impurity effects on the magnetism,
compared with other QSL candidates.
One is that this compound is free from an intermixture of cations
as reported for Ba$_{3}$CuSb$_{2}$O$_{9}$
\cite{Nakatsuji-Science336-2012}
and ZnCu$_{3}$(OH)$_{6}$Cl$_{2}$
\cite{Lee-NatMat6-2007},
and thus one can introduce an impurity into
a specific site in a controlled way.
Another is that a Curie tail is hardly visible
at low temperatures in the title compound
unlike BaCu$_{3}$V$_{2}$O$_{8}$(OH)$_{2}$
\cite{Okamoto-JPSJ78-2009}
and Cu$_{3}$V$_{2}$O$_{7}$(OH)$_{2}$$\cdot$2H$_{2}$O
\cite{Hiroi-JPhysConf145-2009},
in which a substantial contribution of unwanted impurities obscures
the intrinsic physical properties in macroscopic measurements.

Here we report the dilute magnetic impurity effect on
the magnetic properties of Ba$_{3}$ZnRu$_{2}$O$_{9}$
in comparison with Ba$_{3}$CaRu$_{2}$O$_{9}$.
We have found that the magnetic ground state of each compound stands
against 2\% substitution of magnetic impurities
of Cu, Ni, and Co for Zn or Ca,
and these impurities behave as paramagnetic spins.
We have further found that the magnetic response of the paramagnetic spins
depends on the host materials,
which appears in the difference of the Weiss temperature,
which can be hardly explained by
the simple impurity-impurity interaction alone
in the case of the substituted QSL candidate.
We discuss a possible origin in terms of a coupling of
the magnetic impurities with the Ru$^{5+}$ ions,
which would be present only in the substituted Ba$_{3}$ZnRu$_{2}$O$_{9}$.

\section{Experimental}
Polycrystalline samples of Ba$_{3}$\textit{M}Ru$_{2}$O$_{9}$ and
Ba$_{3}$\textit{M}$_{0.98}$\textit{A}$_{0.02}$Ru$_{2}$O$_{9}$
(\textit{M} $=$ Zn and Ca; \textit{A} $=$ Co, Ni, and Cu) were prepared by
solid state reaction using high-purity reagents of
BaCO$_{3}$ (4N), RuO$_{2}$ (3N), ZnO (4N), CaCO$_{3}$ (4N),
Co$_{3}$O$_{4}$ (3N), NiO (3N), and CuO (4N).
Stoichiometric mixtures of the oxides were ground, pressed into pellets,
and pre-sintered in air at 1000$^{\circ}$C for 12 h.
The pre-sintered samples were then re-pelletized after regrinding
and sintered in air at 1200$^{\circ}$C for 72 h.
Powder X-ray diffraction measurements (Cu K$\alpha$ radiation)
at room temperature showed that
all the prepared samples have the hexagonal structure
without any trace of a secondary phase
and the lattice parameters were unchanged by substitution.
The magnetization measurements were conducted by
a Quantum Design superconducting quantum interference device magnetometer.
The magnetic susceptibility ($\chi$) was measured between 2 and 300 K
in an external magnetic field ($H$) of 10 kOe.
The magnetization ($M$) data were collected at 2, 3, and 5 K
in the magnetic field range from 0 to 70 kOe. 
\section{Results and discussion}
\begin{figure}[t]
\centering
\includegraphics[width=80.0mm,clip]{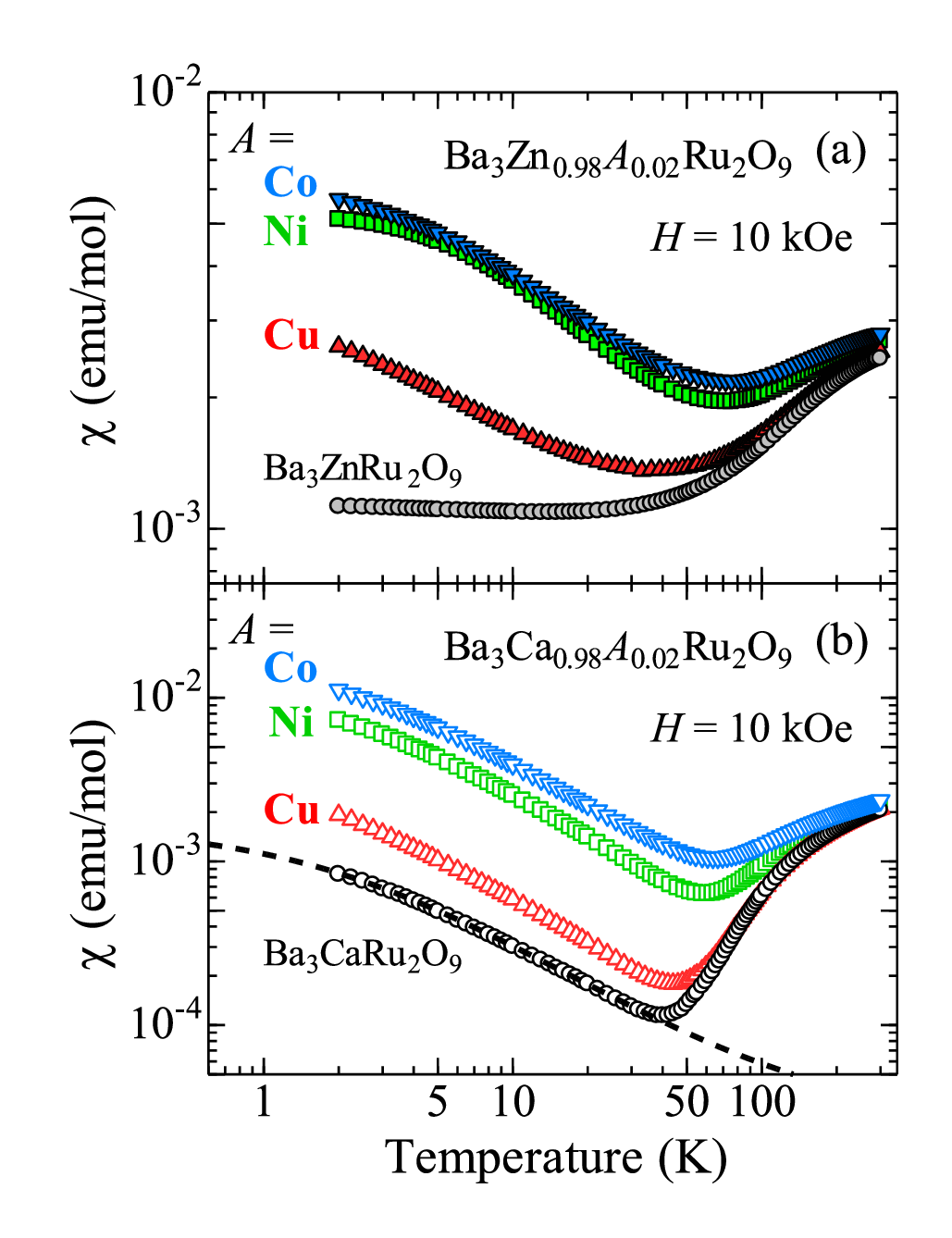}
\caption{(Color online)
Temperature dependence of the magnetic susceptibility measured in 10 kOe for
(a) Ba$_{3}$ZnRu$_{2}$O$_{9}$ and Ba$_{3}$Zn$_{0.98}$\textit{A}$_{0.02}$Ru$_{2}$O$_{9}$,
and (b) Ba$_{3}$CaRu$_{2}$O$_{9}$ and Ba$_{3}$Ca$_{0.98}$\textit{A}$_{0.02}$Ru$_{2}$O$_{9}$
(\textit{A} $=$ Co, Ni, and Cu).
The broken curve in (b) depicts the fit using the Curie-Weiss law (see text).
}
\label{fig:MT}
\end{figure}
Figures \ref{fig:MT}(a) and \ref{fig:MT}(b) show the temperature dependence of
the magnetic susceptibility of Ba$_{3}$\textit{M}Ru$_{2}$O$_{9}$
and Ba$_{3}$\textit{M}$_{0.98}$\textit{A}$_{0.02}$Ru$_{2}$O$_{9}$
(\textit{M} $=$ Zn and Ca; \textit{A} $=$ Co, Ni, and Cu)
on a logarithmic scale.
The susceptibility of Ba$_{3}$ZnRu$_{2}$O$_{9}$ smoothly decreases
with decreasing temperature
and becomes almost constant below 50 K.
The value of $\chi$ at 2 K is equal to about 1.12$\times$ 10$^{-3}$ emu/mol,
which is comparable to those observed in other inorganic spin liquid candidates
\cite{Okamoto-JPSJ78-2009,Hiroi-JPhysConf145-2009}.
Note that no Curie tail is found down to 2 K
as reported in our previous study
\cite{Terasaki-JPSJ86-2017}.
In contrast, $\chi$ does show the Curie tail, i.e.,
it exhibits a rapid increase below 50 K in Ba$_{3}$CaRu$_{2}$O$_{9}$,
which would be due to unwanted magnetic impurities.
We fit this term between 2 and 20 K by assuming the Curie-Weiss law
with a temperature-independent term $\chi_{0}$ given by
$\chi = C/(T + \theta_{\rm W}) + \chi_{0}$,
where $C$ is the Curie constant, $\theta_{\rm W}$ is the Weiss temperature.
The calculated curve reproduces the experimental data
when $C =$ 3.32 $\times$ 10$^{-3}$ emu K/mol,
$\theta_{\rm W} =$ 2.2 K,
and $\chi_{0} =$ 2.5 $\times$ 10$^{-5}$ emu/mol
(figure \ref{fig:MT}(b)).
The obtained value of $C$ corresponds to 0.1\% of
unpaired Ru$^{5+}$ ions out of the total Ru ions.
$\chi_{0}$ is of the order of the Van Vleck paramagnetic susceptibility,
which is consistent with the spin gapped dimer state
of Ba$_{3}$CaRu$_{2}$O$_{9}$\cite{Darriet-JSSC19-1976}.

In all the impurity-substituted samples,
the magnetic susceptibility at around 300 K is
almost the same as that of the corresponding parent compound,
implying that the spin state of the Ru$^{5+}$ ions is
little affected by substitution.
On the other hand, $\chi$ increases significantly below 50 K
as the spin value of the substituent ions increases
in order of Cu$^{2+}$ ($S_{\rm imp} =$ 1/2),
Ni$^{2+}$ ($S_{\rm imp} =$ 1),
and Co$^{2+}$ ($S_{\rm imp} =$ 3/2).
Furthermore, it exhibits a Curie-Weiss-like temperature dependence
without any sign of a magnetic transition down to 2 K.
These results suggest that the Ru$^{5+}$ ions are still responsible for
the ground state of each parent compound,
and the substituted magnetic impurities behave as paramagnetic spins.
Note that the impurity substitution for the \textit{M} sites does not
directly disrupt the network of the Ru$_{2}$O$_{9}$ dimers.

\begin{table}
\centering
\caption{Results of a fit of
$\chi = C_{\rm imp}/(T + \theta_{\rm imp})+ \chi_{\rm Ru}$
to the magnetic susceptibility $\chi$ below 20 K of
Ba$_{3}$\textit{M}$_{0.98}$\textit{A}$_{0.02}$Ru$_{2}$O$_{9}$
(\textit{M} $=$ Zn and Ca; \textit{A} $=$ Co, Ni, and Cu).
$\chi_{\rm Ru}$ is fixed to a specific value for
each of the \textit{M} $=$ Zn and \textit{M} $=$ Ca series
and $x_{\rm imp}$ represents the concentration of
substituted magnetic impurities
estimated from $C_{\rm imp}$ (see text).}
\footnotesize
\begin{tabular}{@{}llll}
\br
Material & $C_{\rm imp}$ (emu K/mol) & $\theta_{\rm imp}$ (K) & $x_{\rm imp}$\\
\mr
(\textit{M}, \textit{A}) $=$ (Zn, Cu) & 8.37$\times$10$^{-3}$ & 3.42 & 0.022\\
(\textit{M}, \textit{A}) $=$ (Zn, Ni) & 4.96$\times$10$^{-2}$ & 9.00 & 0.050\\
(\textit{M}, \textit{A}) $=$ (Zn, Co) & 5.51$\times$10$^{-2}$ & 9.92 & 0.029\\
\mr
(\textit{M}, \textit{A}) $=$ (Ca, Cu) & 6.37$\times$10$^{-3}$ & 1.36 & 0.017\\
(\textit{M}, \textit{A}) $=$ (Ca, Ni) & 3.07$\times$10$^{-2}$ & 2.21 & 0.031\\
(\textit{M}, \textit{A}) $=$ (Ca, Co) & 4.77$\times$10$^{-2}$ & 2.27 & 0.025\\
\br
\end{tabular}\\
\label{table:parameters}
\end{table}
\normalsize
Let us evaluate the low-temerature susceptibility
in the substituted samples.
We assume that this can be divided into two terms,
a contribution from the magnetic impurities
and the Ru$^{5+}$ ions,
i.e., $\chi = \chi_{\rm imp} + \chi_{\rm Ru}$.
The former may follow the Curie-Weiss law as
$\chi_{\rm imp} = C_{\rm imp}/(T + \theta_{\rm imp})$.
We note here that not only the substituted impurities
but also unwanted impurities can contribute to this term,
and it is difficult to distinguish them in the susceptibility data.
Namely, $C_{\rm imp}$ may include both contributions
and $\theta_{\rm imp}$ may reflect
the interaction between all these impurities.
At sufficiently low temperatures below 50 K,
the latter term may be regarded as
the temperature-independent term
in the corresponding parent compound.
Accordingly, we fit the data with
a constant $\chi_{\rm Ru}$ of
1.12 $\times$ 10$^{-3}$ emu/mol
for Ba$_{3}$Zn$_{0.98}$\textit{A}$_{0.02}$Ru$_{2}$O$_{9}$
and 2.5 $\times$ 10$^{-5}$ emu/mol
for Ba$_{3}$Ca$_{0.98}$\textit{A}$_{0.02}$Ru$_{2}$O$_{9}$,
respectively.

The results of the fit are summarized in Table \ref{table:parameters},
together with $x_{\rm imp} = C_{\rm imp}/C_{0}$.
Here $C_{0}$ is the Curie constant expected for
a mole of magnetic impurities with $S_{\rm imp}$ and $g = 2$.
Thus ideally, $x_{\rm imp}$ means the concentration of
the substituted magnetic impurities per formula unit,
and  equals 0.02 in the nominal compositions.
One finds that $x_{\rm imp}$ $\sim$ 0.02
except for (\textit{M}, \textit{A}) $=$ (Zn, Ni).
The reason of the exception of the sample is
to be explored at this stage,
but it may reflect that
a magnetic transition is about to occur
at lower temperatures,
as anticipated from the saturation of $\chi$.
Besides, we notice that the contribution from unwanted magnetic impurities
observed in Ba$_{3}$CaRu$_{2}$O$_{9}$ may not be ignored
for (\textit{M}, \textit{A}) $=$ (Ca, Cu).
If one simply subtracts the contribution from $C_{\rm imp}$,
the corrected value is obtained to be 3.05$\times$10$^{-3}$ emu K/mol,
which corresponds to $x_{\rm imp} \sim$ 0.01.
Such a value is also implied by the fact that
the saturation magnetization of the sample is
estimated to be half of that expected for $x_{\rm imp} =$ 0.02,
while the details are to be clarified
by investigating the compositional dependence.

\begin{figure}[!t]
\centering
\includegraphics[width=75.0mm,clip]{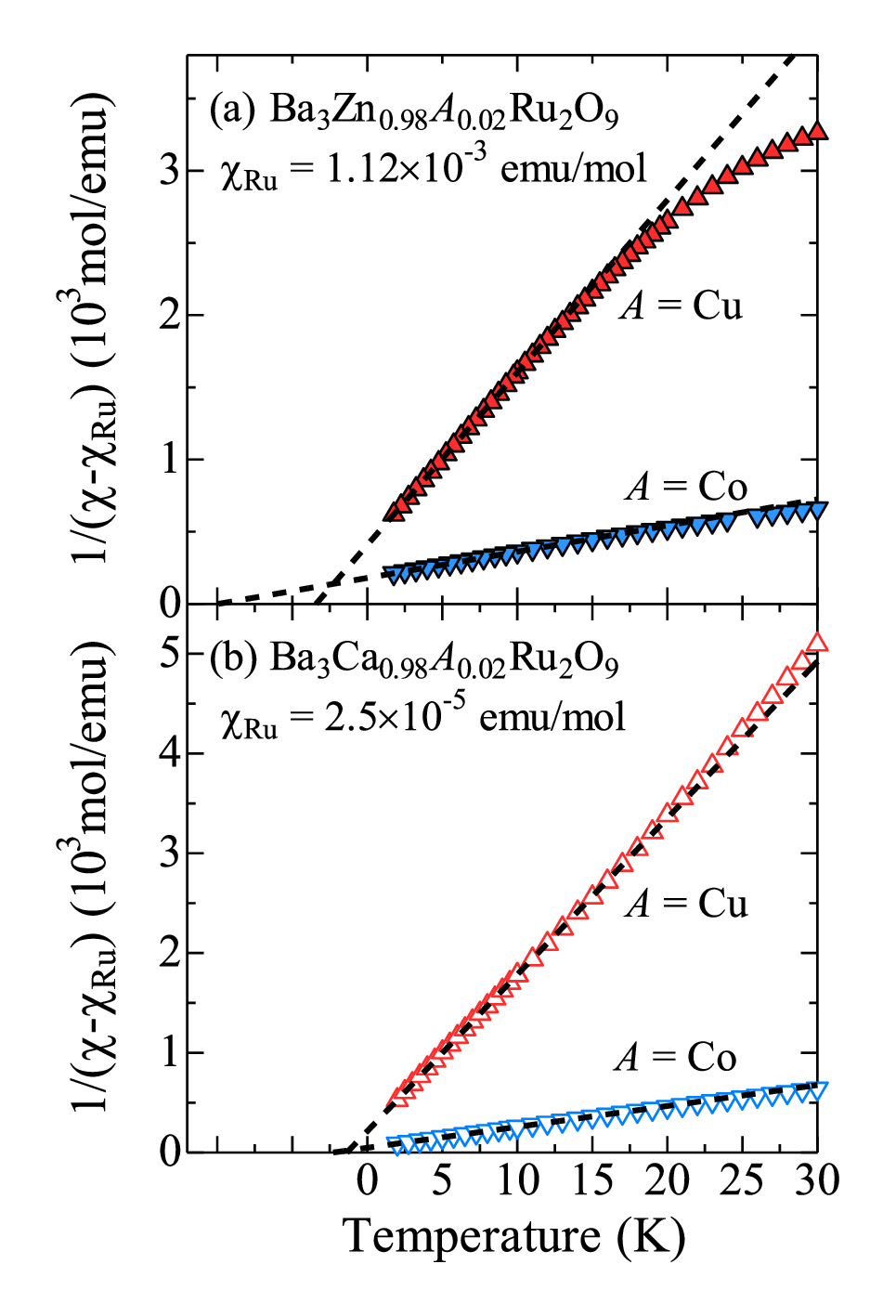}
\caption{(Color online)
Temperature dependence of 1/($\chi - \chi_{\rm Ru}$) below 30 K
for (a) Ba$_{3}$Zn$_{0.98}$\textit{A}$_{0.02}$Ru$_{2}$O$_{9}$
and (b) Ba$_{3}$Ca$_{0.98}$\textit{A}$_{0.02}$Ru$_{2}$O$_{9}$
(\textit{A} $=$ Cu and Co).
The broken lines depict the calculation
using the Curie-Weiss law (see text).}
\label{fig:Curie-Weiss}
\end{figure}

Figure \ref{fig:Curie-Weiss} shows the temperature dependence of
1/($\chi - \chi_{\rm Ru}$) $\simeq$ 1/$\chi_{\rm imp}$  below 30 K of
(a) Ba$_{3}$Zn$_{0.98}$\textit{A}$_{0.02}$Ru$_{2}$O$_{9}$
and (b) Ba$_{3}$Ca$_{0.98}$\textit{A}$_{0.02}$Ru$_{2}$O$_{9}$
(\textit{A} $=$ Cu and Co).
The inverse susceptibility goes to nearly zero
in the limit of $T =$ 0 in the Ca-based materials,
as expected for free spins which obey the Curie law.
On the contrary, 1/$\chi_{\rm imp}$ takes a finite value
in the same limit for the Zn-based materials,
clearly indicating a substantial contribution of the magnetic interaction.
Thus, the magnetic response of the impurity spins seems to
depend on the host materials.
The difference between them can also be seen from the Curie-Weiss fitting,
as depicted in figure \ref{fig:Curie-Weiss} with the broken lines.
In Ba$_{3}$Ca$_{0.98}$\textit{A}$_{0.02}$Ru$_{2}$O$_{9}$,
$\theta_{\rm imp}$ is estimated to be about 1 or 2 K
independently of the magnetic ion in the \textit{A} site,
whereas it increases in the magnitude
with increasing $S_{\rm imp}$
in Ba$_{3}$Zn$_{0.98}$\textit{A}$_{0.02}$Ru$_{2}$O$_{9}$,
from about 3 K for the \textit{A} $=$ Cu sample
to about 10 K for the \textit{A} $=$ Co sample.
These estimated values of $\theta_{\rm imp}$ is anomalously large,
considering that the magnetic interaction between
a few percent of magnetic impurities is
usually found to be of the order of 1 K
in various spin gapped systems and QSL candidates
\cite{Azuma-PRB55-1997,Okamoto-JPSJ78-2009,Bert-PRB76-2007,Lee-PRB96-2017}.
Note that the positive $\theta_{\rm imp}$ implies an antiferromagnetic coupling.

\begin{figure}[!t]
\centering
\includegraphics[width=80.0mm,clip]{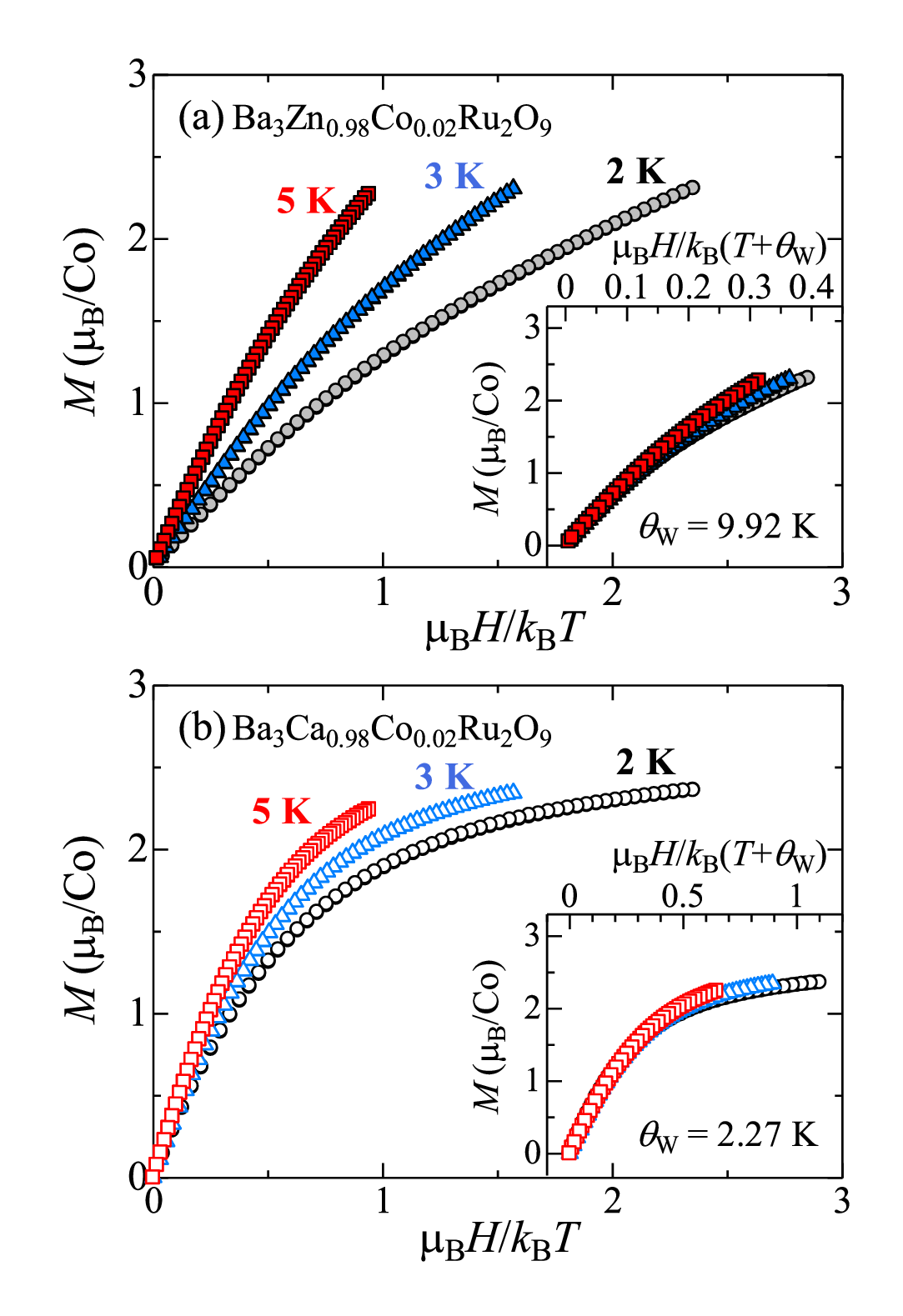}
\caption{(Color online) Magnetization plotted
as a function of $\mu_{\rm B}H/k_{\rm B} T$
at several temperatures
in (a) Ba$_{3}$Zn$_{0.98}$Co$_{0.02}$Ru$_{2}$O$_{9}$
and (b) Ba$_{3}$Zn$_{0.98}$Co$_{0.02}$Ru$_{2}$O$_{9}$.
Each inset represents $M$ versus
$\mu_{\rm B}H/k_{\rm B} (T + \theta_{\rm W}$)
with a certain value of $\theta_{\rm W}$
for the same temperatures.}
\label{fig:MH}
\end{figure}
To further investigate the difference of the magnetic response,
we measure the field dependence of
the magnetization at several temperatures.
For the free spins, the magnetization saturates
for $\mu_{\rm B}H \gg k_{\rm B}T$,
and follows the Brillouin function
as a function of $\mu_{\rm B}H/k_{\rm B}T$.
Here $\mu_{\rm B}$ is the Bohr magneton and
$k_{\rm B}$ is the Boltzmann constant.
Figures \ref{fig:MH}(a) and \ref{fig:MH}(b) show
the magnetization at 2, 3, and 5 K
plotted against $\mu_{\rm B}H/k_{\rm B}T$
for Ba$_{3}$Zn$_{0.98}$Co$_{0.02}$Ru$_{2}$O$_{9}$
and Ba$_{3}$Ca$_{0.98}$Co$_{0.02}$Ru$_{2}$O$_{9}$, respectively.
We find the distinct differences between the two.
In the former sample, the magnetization curves
at each temperature deviate from each other
with increasing magnetic field.
Moreover, $M$ is unlikely to saturate
even at 2 K in 70 kOe.
In contrast, all the experimental data are about to
fall into a single curve with a sign of saturation
in the latter sample.

The slight deviation observed in
Ba$_{3}$Ca$_{0.98}$Co$_{0.02}$Ru$_{2}$O$_{9}$
can be explained by considering the impurity-impurity interaction.
When the paramagnetic spins weakly interact with each other,
the Brillouin function is modified,
and then $M$ scales to $\mu_{\rm B}H/k_{\rm B}(T + \theta_{\rm W})$
\cite{Bert-PRB76-2007,Lee-PRB96-2017},
which means the reduction of the field effect.
As shown in the inset of figure \ref{fig:MH}(b),
a good scaling is indeed established with a small $\theta_{\rm W}$,
which is adopted the same value as $\theta_{\rm imp}$.
This fact allows us to attribute $\theta_{\rm imp}$
in Ba$_{3}$Ca$_{0.98}$\textit{A}$_{0.02}$Ru$_{2}$O$_{9}$
to the weak interaction between the dilute magnetic impurities.
We attempt the same analysis on the magnetization curves
in Ba$_{3}$Zn$_{0.98}$Co$_{0.02}$Ru$_{2}$O$_{9}$
(the inset of figure \ref{fig:MH}(a)).
They overlap each other at low magnetic fields,
but the scaling gets worse as $H$ increases,
implying that the magnetic response is affected
by other contributions.
A similar trend is found for the Cu-substituted samples (not shown).

Now let us discuss a possible origin of
the relatively larger Weiss temperature found
in Ba$_{3}$Zn$_{0.98}$\textit{A}$_{0.02}$Ru$_{2}$O$_{9}$.
The difference in the magnetic response
suggests that the substituted magnetic impurities
in each host lattice exist
under different environments.
This would be related to the magnetism of the Ru$^{5+}$ ions:
the impurities can interact with the Ru$^{5+}$ ions
only in the Zn-based compounds.
In this context, it is worth noting that
the low $T$ upturn of the susceptibility
due to the impurities is
strongly suppressed below $\theta_{\rm imp}$
in Ba$_{3}$Zn$_{0.98}$\textit{A}$_{0.02}$Ru$_{2}$O$_{9}$.
This feature is reminiscent of the low-temperature
susceptibility in Kondo systems
\cite{Sumiyama-JPSJ55-1986,Marumoto-PRB54-1996},
in which localized spins are screened by conduction electrons
via an antiferromagnetic coupling between them.
Therefore, we propose that the large Weiss temperature
observed in the substituted Ba$_{3}$ZnRu$_{2}$O$_{9}$ is
the implication of such screening effect on the impurity spins,
which results from the magnetic coupling of these spins and the Ru$^{5+}$ spins.
If that were the case, the spin fluctuation rates of
impurity spins should be enhanced,
leading to the magnetic moment instability
\cite{Khatua-PRL68-1992}.
In this sense, the Weiss temperature is interpreted as
a measure of the degree of the spin fluctuations.
The screening effect would also be suggested from
the absence of a Curie tail in the parent Ba$_{3}$ZnRu$_{2}$O$_{9}$
despite unwanted magnetic impurities should exist,
as in Ba$_{3}$CaRu$_{2}$O$_{9}$.

Unlike Kondo systems, there are no conduction electrons in the title compound.
Nevertheless, QSL candidates are believed to possess
a kind of quasiparticles as fermionic elementary excitations,
because a gapless $T$-linear term of
the specific heat has been generally observed
\cite{Okamoto-JPSJ78-2009,Terasaki-JPSJ86-2017, Helton-PRL98-2007,Yamashita-Nature4-2008}.
Thus, one possible explanation of the screening effect is
that a magnetic impurity couples with the quasiparticles.
However, this scenario is incompatible with
the $S_{\rm imp}$ dependence of $\theta_{\rm imp}$ in our system
because the Weiss temperature exhibits the opposite $S_{\rm imp}$ dependence
in the framework of the Kondo model
\cite{Gruner-Advances23-1974}.
As another possible scenario,
we suggest that the enhancement of
the spin fluctuation rates is caused by
the quantum fluctuations originating from
the Ru$_{2}$O$_{9}$ dimers.
In our previous study
\cite{Terasaki-JPSJ86-2017},
we have pointed out that
the spin state of the Ru$_{2}$O$_{9}$ dimers dynamically changes
in the range between $S_{\rm tot} = 0$ and $S_{\rm tot} = 2$
for the title compound,
where $S_{\rm tot}$ is the total spin.
In such a situation, a magnetic coupling of the magnetic impurities
with the  Ru$_{2}$O$_{9}$ dimers will be dynamical
and produces a fluctuating internal field
at each impurity site.
The increase in $S_{\rm imp}$ would imply
a stronger exchange coupling between them,
resulting in the observed $S_{\rm imp}$ dependence of $\theta_{\rm imp}$. 
The NMR and neutron scattering measurements are indispensable
to examine this scenario through direct observation of
the spin fluctuation.

\section{Summary}
We have investigated the effect of
the magnetic impurity on the magnetic properties
of the quantum spin liquid candidate
Ba$_{3}$ZnRu$_{2}$O$_{9}$
through the comparison with
the spin gapped compound Ba$_{3}$CaRu$_{2}$O$_{9}$.
The magnetic ground state of these systems is robust
against the 2\% impurity substitution,
and the introduced magnetic impurities
behave as paramagnetic spins.
We have found that the magnetic response of these paramagnetic spins
is dependent on the host materials
and the difference between the two manifest itself
in the Weiss temperature.
We have proposed a possible picture that
the enhanced Weiss temperature observed
in the substituted QSL candidate is
the implication of a kind of screening effect
on the impurity spins,
which arises from a dynamical magnetic coupling between
the magnetic impurities and the Ru$_{2}$O$_{9}$ dimers.
%
\ack{
We gratefully acknowledge Y. Hara, T Matsushita,
and N. Wada at Nagoya University
for the collaboration in the magnetic and
thermodynamic measurements at very low temperatures.
This work was partly supported by Grants-in-Aid for
Scientific Research (18H01173).
One of the authors (T. D. Y.) was supported by the Program for Leading Graduate
Schools "Integrative Graduate Education and Research in Green Natural Sciences",
MEXT, Japan and a Grant-in-Aid for JSPS Research Fellow (No. 17J04840), MEXT, Japan.}
%
%

\end{document}